\documentstyle[11pt,aaspp4]{article}

\received{January 27 1997}
\accepted{March 31 1997}

\slugcomment{To be published in the Astrophysical Journal (Letters) June 20
1997 issue, Volume 482}

\lefthead{Wilson, Binette \& Storchi-Bergmann}
\righthead{Extended Gas in Active Galaxies}

\begin{document}

\title{The Temperature of Extended Gas in Active
Galaxies -- Evidence for Matter-Bounded Clouds}

\author{A. S. Wilson\altaffilmark{1,3}}
\affil{Space Telescope Science Institute, 3700 San Martin Drive, Baltimore, MD
21218}

\author{L. Binette\altaffilmark{2}}
\affil{Instituto de Fisica, UFRGS, Campus do Vale, CxP 15051, 91501-970
Porto Alegre - RS, Brazil}

\and

\author{T. Storchi-Bergmann\altaffilmark{3}}
\affil{Instituto de Fisica, UFRGS, Campus do Vale, CxP 15051, 91501-970
Porto Alegre - RS, Brazil}


\altaffiltext{1}{Also Astronomy Department, University of Maryland,
College Park, MD 20742}
\altaffiltext{2}{Also European Southern Observatory, 
Casilla 19001, Santiago 19, Chile}
\altaffiltext{3}{Visiting Astronomer, Cerro Tololo Inter-American
Observatory, operated by the Association of Universities for Research in
Astronomy, Inc., under contract with the National Science Foundation}


\begin{abstract}

We report measurements of the electron temperature at about a dozen locations
in the extended emission-line regions of five active (Seyfert and radio)
galaxies. Temperatures (T$_{[OIII]}$ and T$_{[NII]}$) have been determined from
both the I([OIII]$\lambda$4363)/I([OIII]$\lambda$5007) and
I([NII]$\lambda$5755)/I[NII]$\lambda$6583) ratios. 
T$_{[OIII]}$ lies in the
range (1.0 -- 1.7) $\times$ 10$^{4}$K. We find a strong trend for
T$_{[OIII]}$ to be higher than T$_{[NII]}$, with the difference typically being
$\approx$ 5,000K. 
Because the critical density for collisional de-excitation
of the $^{1}$D${_2}$ level in NII is lower than that of the same level
in OIII, the deviations of the measured intensity ratios from those
expected for T$_{[OIII]}$ = T$_{[NII]}$ in the low density limit are unlikely
to result from
collisional de-excitation.
The measured values of T$_{[OIII]}$ and the
differences between T$_{[OIII]}$ and T$_{[NII]}$ are very similar to those
found in Galactic planetary nebulae. It is argued that the dominant form
of energy input to the clouds is photoionization, but detailed modelling 
indicates that the temperature difference is too large to be accounted for 
in terms of
photoionization of ionization-bounded clouds. We propose instead that both
matter- and ionization-bounded clouds are present in the extended emission-line
regions, with most of the [OIII] emission originating from a hot zone
in the matter-bounded clouds and essentially all of the [NII] from the 
ionization-bounded clouds.

\end{abstract}




%

\section{Introduction}

High ionization, narrow emission lines 
are a defining characteristic of the
nuclei of Seyfert
galaxies and are often seen in radio galaxies. The gas emitting these lines
is generally believed to be photoionized, though the origin of the ionizing
photons is still debated.
Spectroscopic studies have
focussed mostly on the spatially unresolved nuclei, the narrow line
emission of which originates from gas with a wide range of densities
(10$^{2}$ $\le$ n$_{e}$ $\le$ 10$^{7}$ cm$^{-3}$). This large spread of
densities, together with the unknown distances of the individual gas
clouds from the source of ionizing photons, leads to serious ambiguities
in photoionization models. For example, collisional de-excitation of the
$^{1}$D${_2}$ level of OIII becomes significant for
n$_{e}$ $\ge$ 10$^{5}$ cm$^{-3}$, rendering the intensity ratio
R$_{[OIII]}$ = I$(\lambda4363)$/I$(\lambda5007)$ an unreliable measure
of electron temperature (Osterbrock 1989). 

The high ionization gas is, however,
often spatially extended in ground-based observations of both Seyfert
(e.g. Mulchaey, Wilson \& Tsvetanov 1996) and radio (e.g. Baum \&
Heckman 1989) galaxies. Study of such extended gas provides two advantages.
First, the density is low (n$_{e}$ $\le$ 10$^{3}$ cm$^{-3}$ -- e.g. 
Morganti et al. 1991; Tadhunter et al. 1994), so that collisional
deexcitation is unimportant for most forbidden lines, the exceptions
providing density diagnostics. Second, the geometric properties of
the gas, including its distance from the nucleus, can be measured directly.
These extended emission-line regions (EELRs) thus provide much simpler physical
situations than the spatially unresolved narrow or broad line regions, and
their excitation should be correspondingly easier to understand. 

Tadhunter, Robinson \& Morganti (1989) have presented measurements of
R$_{[OIII]}$ for twelve locations in EELRs around six
active galaxies. These measurements imply electron temperatures in the
range 12,800 $<$ T$_{[OIII]}$ $<$ 22,000K, whereas photoionization models
which successfully account for most other line ratios 
predict T$_{[OIII]}$ $<$ 11,000K. These high
measured temperatures were confirmed by Storchi-Bergmann et al. (1996,
hereafter SBWMB),
who found 10,000 $<$ T$_{[OIII]}$ $<$ 17,000K for EELRs around five
active (both Seyfert and radio) galaxies.
Tadhunter, Robinson \& Morganti (1989) concluded that
either the EELRs have a heating source in addition to photoionization --
plausibly cosmic rays or shocks -- or the metal abundances are lower than
the solar values assumed in the models. Binette, Wilson \& Storchi-Bergmann
(1996, hereafter BWSB) favored an alternative picture 
in which there are two populations
of ionized clouds -- a matter-bounded (MB) component responsible for
most of the HeII and high-ionization forbidden line emission and an
ionization-bounded (IB) component emitting low- to intermediate-ionization
forbidden lines. In this model, the IB clouds are illuminated by an ionizing
continuum modified by absorption in the MB clouds. 
A new sequence of photoionization models was
thereby obtained by varying the ratio, A$_{M/I}$, 
of the solid angle subtended by
the MB clouds at the ionizing source to that subtended by the
IB clouds. The main success of the A$_{M/I}$ sequence of models is that it
provides a natural explanation for observed correlations between both
the I([NeV]$\lambda$3426)/I([OII]$\lambda\lambda$3727) and 
I([OIII]$\lambda$5007)/I([OII]$\lambda\lambda$3727) ratios and the 
I(HeII$\lambda$4686)/I(H$\beta$) ratio (BWSB). Such 
a correlation between the gaseous
ionization, measured from the forbidden lines,
and the I(HeII$\lambda$4686)/I(H$\beta$) ratio cannot be understood
in terms of the standard $U$ (ionization parameter) sequence, since
the HeII$\lambda$4686/H$\beta$ ratio for IB clouds depends primarily on
the slope of the ionizing continuum between 13.6 and 54.4 eV, and is
relatively insensitive to $U$. Further, the
electron temperature T$_{[OIII]}$ is
correctly predicted by the A$_{M/I}$ sequence,
provided the thickness and ionization parameter of the MB component
are appropriately selected. Finally, much stronger high ionization lines
(e.g. [NeV]$\lambda$3426 and CIV$\lambda\lambda$1549) are expected in
the A$_{M/I}$ sequence than in the $U$ sequence, again in accord with 
observations.

In the A$_{M/I}$ sequence, the [OIII]$\lambda$5007 and $\lambda$4363 emission
originates
mostly in the MB clouds, while essentially all of [NII]$\lambda$6583 and
$\lambda$5755 emission comes from the IB clouds. Because the MB clouds are
hotter than the IB ones (see Figs 1 and 2 of BWSB),
the temperature T$_{[OIII]}$ inferred from R$_{[OIII]}$ should be
considerably larger than the temperature T$_{[NII]}$ inferred from
R$_{[NII]}$ = I($\lambda5755)$/I($\lambda6583)$. BWSB predict that T$_{[OIII]}$
should be $\simeq$ 5,000K larger than T$_{[NII]}$ in the 
A$_{M/I}$ sequence, but only slightly higher (by $<$ 1,000K) in the
$U$-sequence. The temperature T$_{[NII]}$ is predicted to be 9,200K
for solar abundances in the
A$_{M/I}$ sequence, given the value of $U$ assumed by BWSB for the
IB clouds.
These predictions of the A$_{M/I}$ sequence models are in good
agreement with temperature measurements of the NW cloud of Cygnus A,
which give T$_{[OIII]}$ = 15,000 $\pm$ 1,000K and
T$_{[NII]}$ = 10,000 $\pm$ 600K (Tadhunter, Metz \& Robinson 1994). 

In a typical Seyfert galaxy,
I([OIII]$\lambda$5007)/I(H$\beta$) $\simeq$ 10,
I(H$\alpha$)/I(H$\beta$) $\simeq$ 3
and I([NII]$\lambda$6583)/I(H$\alpha$) $\simeq$ 1. For T$_{e}$ = 10$^{4}$K,
R$_{[NII]}$ = 0.016 and R$_{[OIII]}$ = 0.0064 (Osterbrock 1989). Thus,
if T$_{[NII]}$ = T$_{[OIII]}$ = 10$^{4}$K, [NII]$\lambda$5755 and
[OIII]$\lambda$4363 should have roughly the same strength
(I([OIII]$\lambda$4363) $\simeq$ 1.3 $\times$ I([NII]$\lambda$5755) and be about
equally easy to detect with modern CCD spectrographs. However, this
expectation is not borne out by our observations. 

The
purpose of the present letter is to provide further measurements of 
T$_{[NII]}$ in EELRs and evaluate the implications for the gaseous excitation.
We find a strong tendency for T$_{[OIII]}$ to exceed
T$_{[NII]}$ by several thousand degrees and note that a similar trend, with
a similar range of temperatures, is found in Galactic planetary nebulae.
It is argued that the most likely explanation is that the EELRs contain both
matter- and ionization-bounded clouds.

\section{Results}

In their study of nuclear and EELR spectra of 5 active galaxies, 
SBWMB report a detection of [NII]$\lambda$5755 at 
only one location - the nucleus of Mkn 573. We have, therefore, measured upper
limits to the flux of this line at the 12 other locations where
SBWMB obtained spectra.
Two methods were used. In the first method, a spectrally unresolved
artificial line was superimposed on the continuum and its flux adjusted by
eye to provide an upper limit to the actual line flux. In the second method,
the formal r.m.s. noise in the continuum was calculated and the 3 $\times$
r.m.s. upper limit on the flux of an unresolved spectral line was determined.
In some cases, the upper limits for these two approaches agreed quite well,
while in others the first method gave a higher value. We decided to adopt the
more conservative upper limits from the first method.

One concern was that the stellar template spectrum (which was subtracted from
the observed spectrum prior to measuring the emission lines - see
SBWMB) might, for some unknown reason, be a poor
match near $\lambda$5755 rest wavelength. If the template were too high, a real
emission line could have been mistakenly subtracted away. Examination of
of the two templates used by SBWMB showed them to be in excellent agreement
in the spectral region in question and that $\lambda$5755 corresponds to a
local minimum. Since the templates match the stellar continuum very well
almost everywhere else, we have no reason to doubt the
validity of the template
subtraction.

The resulting upper limits to [NII]$\lambda$5755,
and the fluxes of [NII]$\lambda$6583, [OIII]$\lambda$4363 and
[OIII]$\lambda$5007, were
corrected for obscuration, when present, using SBWMB's
values of A$_{V}$,
which were obtained assuming an intrinsic
emission-line Balmer decrement
of H$\alpha$/H$\beta$ = 3.1. Correction for obscuration was not possible for
the nucleus of ESO 362-G8, because a reliable measurement of H$\beta$
could not be obtained due to the A type stellar continuum; the nucleus of
this galaxy is omitted from the following discussion.
The measurements of or limits to
R$_{[NII]}$ and R$_{[OIII]}$ were then converted to T$_{[NII]}$ and
T$_{[OIII]}$ using equations 5.5 and 5.4 
of Osterbrock (1989), respectively, assuming
the low density limit. 

The results are shown in the left hand panel of Fig. 1, in which T$_{[OIII]}$
is plotted against T$_{[NII]}$. It is immediately apparent that T$_{[OIII]}$
tends to be greater than T$_{[NII]}$ by an amount which ranges up to
$\approx$ 7,000K, but
is more typically $\approx$ 5,000K. This trend is found for the nucleus of
NGC 526A and for most of the off-nuclear locations
(only upper limits to both temperatures are available for 
the nuclei of PKS 0349-278 and PKS 0624-206). It is unlikely that this
trend is related to collisional de-excitation of the $^{1}$D$_{2}$ levels,
since a) the critical densities of 8.6 $\times$ 10$^{4}$ cm$^{-3}$ (for
NII) and 7.0 $\times$ 10$^{5}$ cm$^{-3}$ (for OIII) are much higher than
expected for off-nuclear gas, and b) such collisional de-excitation
effects would tend to depress
[NII]$\lambda$6583 at a lower density than [OIII]$\lambda$5007, thus 
{\it increasing} R$_{[NII]}$ and the calculated value of T$_{[NII]}$ 
relative to R$_{[OIII]}$ and T$_{[OIII]}$.

The right panel of Fig. 1 shows a similar plot for Galactic planetary nebulae
using the measurements of Kingsburgh \& Barlow (1994). The range of [OIII]
temperatures is similar to that seen in the AGN and there is also a trend
for T$_{[OIII]}$ to exceed T$_{[NII]}$
(see also Torres-Peimbert \& Peimbert 1977; Kaler 1986). The trends between
T$_{[OIII]}$ and T$_{[NII]}$ in planetary nebulae are generally considered to be
excitation effects in a photoionized gas (Kaler 1986). The similar behavior
of the AGN EELRs then strongly suggests that the same may be true for them.

\section{Discussion}

The photoionization code MAPPINGS IC (Binette et al. 1993; Ferruit et al. 1997)
has been used to calculate the expected values of R$_{[OIII]}$ and
R$_{[NII]}$ in various situations. We began by considering IB,
isobaric, dust-free
($\mu = 0$)\footnote{$\mu$ is the dust-to-gas ratio of the cloud in units
of the solar neighborhood dust-to-gas ratio.}
clouds of solar metallicity (Z = 1) and low density 
($n$ = 1,000 cm$^{-3}$ at the irradiated face)
photoionized
by continua of various shapes -- 
blackbodies, power laws and a spectrum with a uv bump. A grid of models was
generated by varying
the ionization parameter $U$. The results are plotted in Fig. 2, 
in which the models represented by the long dashed (power-law ionizing spectrum)
and continuous (blackbody-shaped ionizing spectrum)
lines have
[OIII]$\lambda$5007/H$\beta$
$\le$ 16, and are thus consistent with the observed values of this ratio,
while the dotted extensions have [OIII]$\lambda$5007/H$\beta$
$>$ 16, and are thus unacceptable. The models with
[OIII]$\lambda$5007/H$\beta$ $\le$ 16
do show deviations from the 
T$_{[OIII]}$ = T$_{[NII]}$ line, but these differences between
T$_{[OIII]}$ and T$_{[NII]}$ are much smaller than are observed.
A power-law model with very low metallicity
($Z = 0.2$) provides [OIII] temperatures in general
agreement with those observed, even without dust ($\mu = 0$),
but again the difference between 
T$_{[OIII]}$ and T$_{[NII]}$ is too small. Also, such a low metallicity is
implausible for the circumnuclear emission line regions discussed here.
A high density
($n$ = 60,000 cm$^{-3}$) model lies to the right of the
T$_{[OIII]}$ = T$_{[NII]}$ line, confirming that collisional de-excitation
tends to {\it increase} R$_{[NII]}$ relative to R$_{[OIII]}$, which is opposite
to the trend observed.

As a second step, we have considered MB clouds. We adopt the model of BWSB,
which contains two cloud populations -- high ionization MB clouds and low
ionization IB clouds.
The two vertical
dashed lines in Fig. 2 show the predicted behavior of R$_{[OIII]}$ and
R$_{[NII]}$ as a function
of the parameter A$_{M/I}$ (defined in Section 1). 
The model marked `A$_{M/I}$ $\alpha$ = --1.3' is the same
as described in BWSB, and represents a varying proportion of MB
clouds (with $U_{MB}$ = 0.04) and IB clouds (with
$U_{IB}$ = 5.2 $\times$ 10$^{-4}$). The model marked 
`A$_{M/I}$ $\alpha$ = --1.1' has a harder continuum, $U_{MB}$ = 0.03 and
U$_{IB}$ = 1.8 $\times$ 10$^{-4}$. The
predictions of
these models are in agreement
with the measured values in the nucleus of Mkn 573 and the NW cloud
of Cygnus A and are consistent with most of the other points. 
These models were also shown by BWSB to be in agreement with the other
optical line ratios. It should be emphasised that
our models assume a single
ionization parameter and total column
for each of the matter- and ionization-bounded populations of 
clouds, which represents a gross oversimplification of a probably
complex situation involving a continuous range of
$U_{MB}$, $U_{IB}$ and the column density through the clouds.
The model predictions should thus be
considered as illustrative rather than definitive. Nevertheless, the key
prediction (BWSB) of the matter- plus ionization-bounded
model -- that T$_{[OIII]}$ should exceed T$_{[NII]}$ by 
$\approx$ 5,000K -- seems to be borne out by the data.

Potential sources of uncertainty are non-collisional contributions
to [OIII]$\lambda$4363. Such contributors include charge transfer
(e.g. Dalgarno \& Sternberg 1982) and
recombination (e.g. Rubin 1986). 
The relative contribution of these effects is lower
at higher temperatures because of 
the exponential temperature dependence of  
collisional excitation. Although no detailed calculations are available
in the AGN context, and population of the $^{1}$S$_{0}$ level of OIII by
these processes is not
included in MAPPINGS IC, we infer from calculations with hot
stellar ionizing sources
(Kingdon \& Ferland 1995) that the resulting error is less than 300K
when T$_{[OIII]}$ $>$ 10,000K. Therefore, recombination processes make a
very minor contribution to [OIII]$\lambda$4363 in the AGN context.

Lastly, there is the question of whether sources of energy other than
photoionization may contribute to the temperature difference between
[OIII] and [NII]. Such sources include shocks, which may be generated
by outflowing winds or jets, and relativistic particles.
While temperature stratification is expected in classical shock
models (e.g. Binette, Dopita \& Tuohy 1985), it
is important to emphasise that the line ratios seen in the extended
emission line regions under study are characteristic of 
photoionization by a hard spectrum (BWSB).
Shocks with photoionized precursors can 
generate spectra which resemble those of Seyfert galaxies (Dopita \& Sutherland
1995), but then the temperature is expected to be similar to that found
in classical photoionization models of IB clouds
(see Fig. 7 of Dopita \& Sutherland 
1995). Relativistic particles may contribute heating (Ferland \& Mushotzky
1984), but most of the off-nuclear 
regions we have 
studied do not have strong radio emission. Also, it is unclear why
the relativistic particles would heat the
OIII region preferentially over the NII.

In summary, the range of electron temperature 
and the difference between T$_{[OIII]}$
and T$_{[NII]}$ are very similar in planetary nebulae and the EELRs of
active galaxies. These similarities do not favor a major contribution 
to the temperature from
shock heating in the galaxies, despite the higher gas velocities there.
More generally, the similarities between the two classes of object
support the idea that the difference between the [OIII] and [NII]
temperatures is primarily related to the 
structure of the ionized clouds, rather than the presence of
extra heating sources or a property of the ionizing source. Indeed,
the presence of matter-bounded clouds in some planetary nebulae is
demonstrated by differences between the two Zanstra temperatures derived
from the nebular H I and He II lines (Osterbrock 1989, p 150), and also
possibly by discrepancies between the Zanstra temperatures 
and the spectroscopic effective temperature of
the central star (e.g. Kudritzki \& M\'endez 1989).

ASW thanks J. P. Harrington for several valuable discussions. This research
was supported in part by NASA under grants NAGW 4700 and NAG 81027 and
by the Space Telescope Science Institute through grants GO5411 and GO6006.
LB is grateful to CNPq and CAPES of Brazil for financial assistance.

\vfil\eject

\centerline{\bf CAPTIONS TO FIGURES}
\medskip
\noindent
Figure 1 --- Left: A plot of [OIII] temperature versus [NII] temperature
(in K) for nuclear and extended gas in active galaxies. Temperatures are plotted
for all locations at which spectra were obtained by Storchi-Bergmann
et al. (1996),
except for the nucleus of ESO 362-G8 (see text) and
10$^{\prime\prime}$ NW of the nucleus of PKS 0634-206, at which only a poor
upper limit on T$_{[NII]}$ could be obtained.
The error bars are dominated by the uncertainties in the fluxes of
[OIII]$\lambda$4363 and [NII]$\lambda$5755.
The straight line represents T$_{[OIII]}$ = T$_{[NII]}$.
The temperature sensitive [NII]$\lambda$5755 line is detected in only the
nucleus of Mkn 573, so the [NII] temperatures are upper limits at all other
positions. Right: A similar plot for planetary nebulae, constructed
from the temperatures and their errors
listed by Kingsburgh \& Barlow (1994). It is notable that the range of
T$_{[OIII]}$ is similar in the two types of object and that there is a
trend for T$_{[OIII]}$ $>$ T$_{[NII]}$ in both types.

\medskip
\noindent
Figure 2 --- A plot of R$_{[OIII]}$ = I$(\lambda4363)$/I$(\lambda5007)$ versus
R$_{[NII]}$ = I($\lambda5755)$/I($\lambda6583)$. The diamonds represent
the measured ratios in the nuclei of Mkn 573 and the NW cloud of Cygnus A
(Tadhunter, Metz \& Robinson 1994), while the error bars
and upper limits represent
our other measurements (see Fig. 1). The straight dot--dashed line is
the locus T$_{[OIII]}$ = T$_{[NII]}$ in the low density limit; the
three asterisks represent temperatures of 8,500K, 10,000K and 15,000K
(from lower left to upper right). All other lines
represent model predictions calculated with
MAPPINGS IC. Unless otherwise
noted, the models are for ionization-bounded, dust-free, isobaric clouds of
solar abundance and density 1,000 cm$^{-3}$. The two continuous
lines are sequences of ionization parameter
with a black-body shaped
ionizing continuum having the
temperatures indicated. For these models represented by the
continuous lines, the ratio
[OIII]$\lambda$5007/H$\beta$ $\le$ 16, in accord with observations. The dotted
extensions to the continuous lines have [OIII]$\lambda$5007/H$\beta$
$>$ 16. The four long dashed lines
are sequences of ionization parameter for ionization-bounded clouds,
with indicated abundance (Z) and dust content ($\mu$), ionized by
a power-law shaped ionizing continuum of index $\alpha$ = --1.3. These
four models have [OIII]$\lambda$5007/H$\beta$ $\le$ 16 where the line is
long dashes, but [OIII]$\lambda$5007/H$\beta$
$>$ 16 for the dotted extensions.
The long dashed curve to the right is for high density (60,000 cm$^{-3}$) 
clouds.
The long dash -- short dash line towards
the bottom is a sequence of ionization parameter
for a broken power law ionizing spectrum (simulating a uv bump), following
Mathews \& Ferland (1987).
Lastly, the two vertical short dashed lines
to the left are models of the $A_{M/I}$ sequence with Z = 1, $\mu$ = 0.015
and power-law
ionizing continua of index $\alpha$ = --1.3 and $\alpha$ = --1.1. The parameter
$A_{M/I}$ increases from 0.04 at the bottom to 16 at the
top.

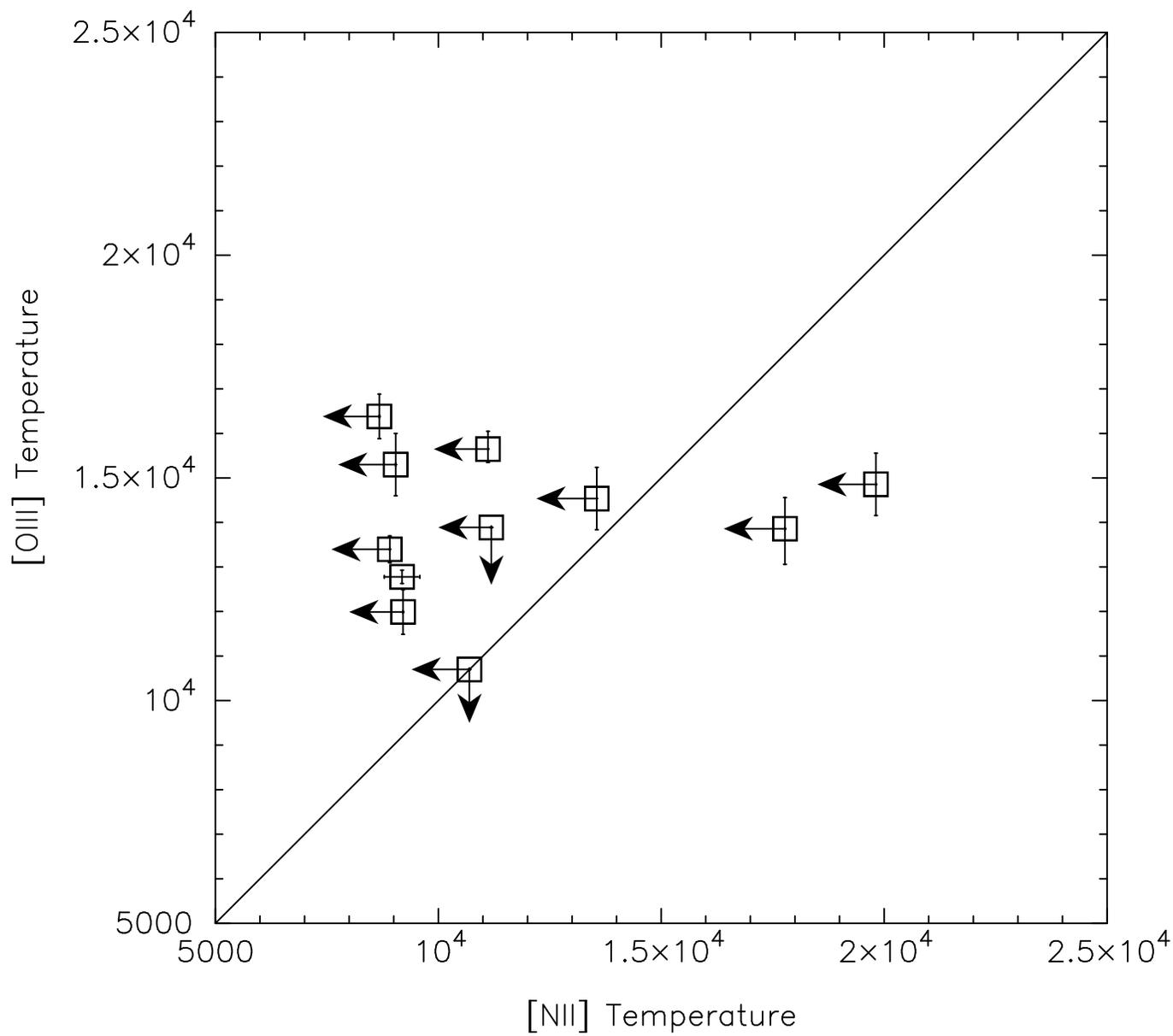
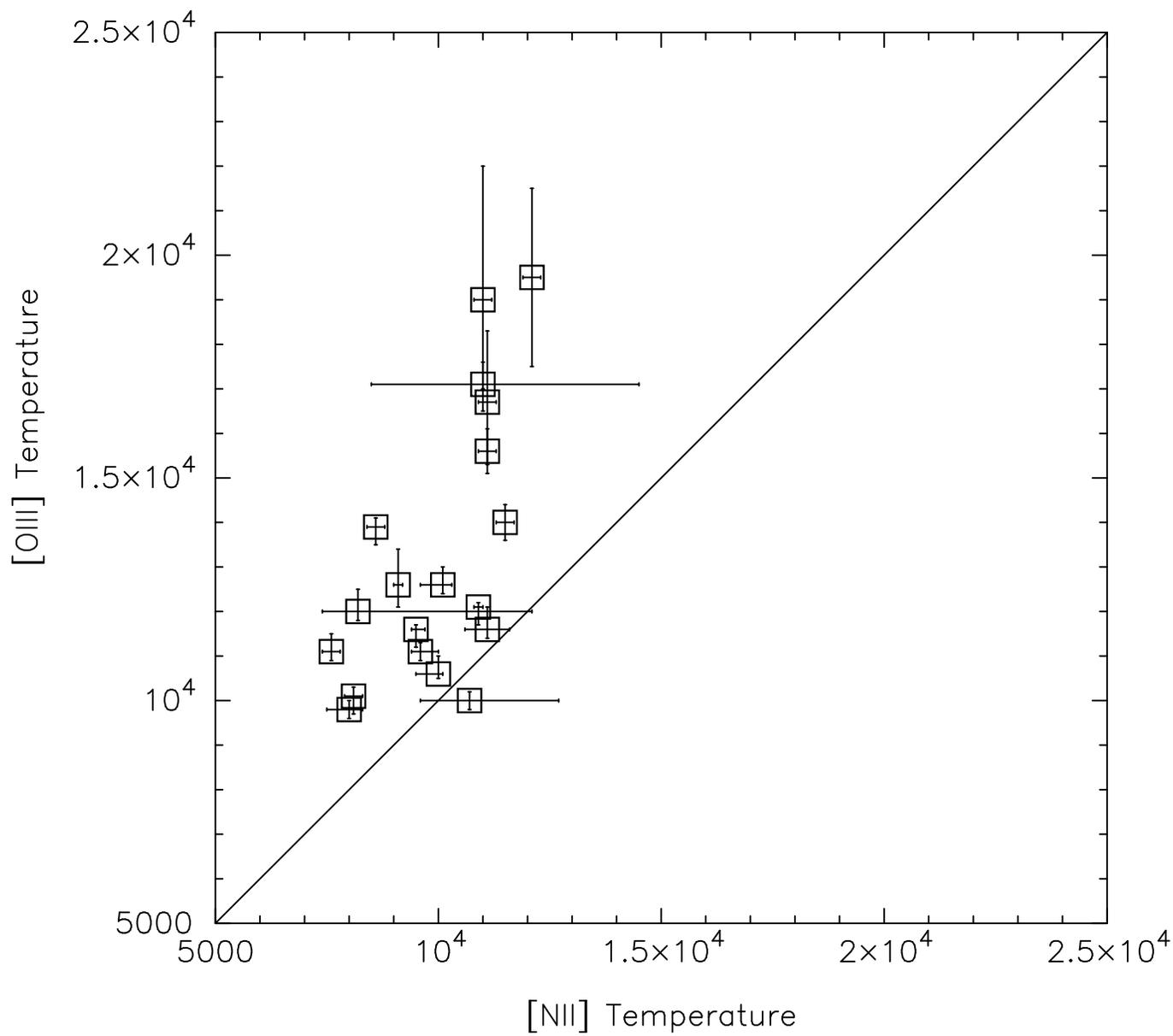
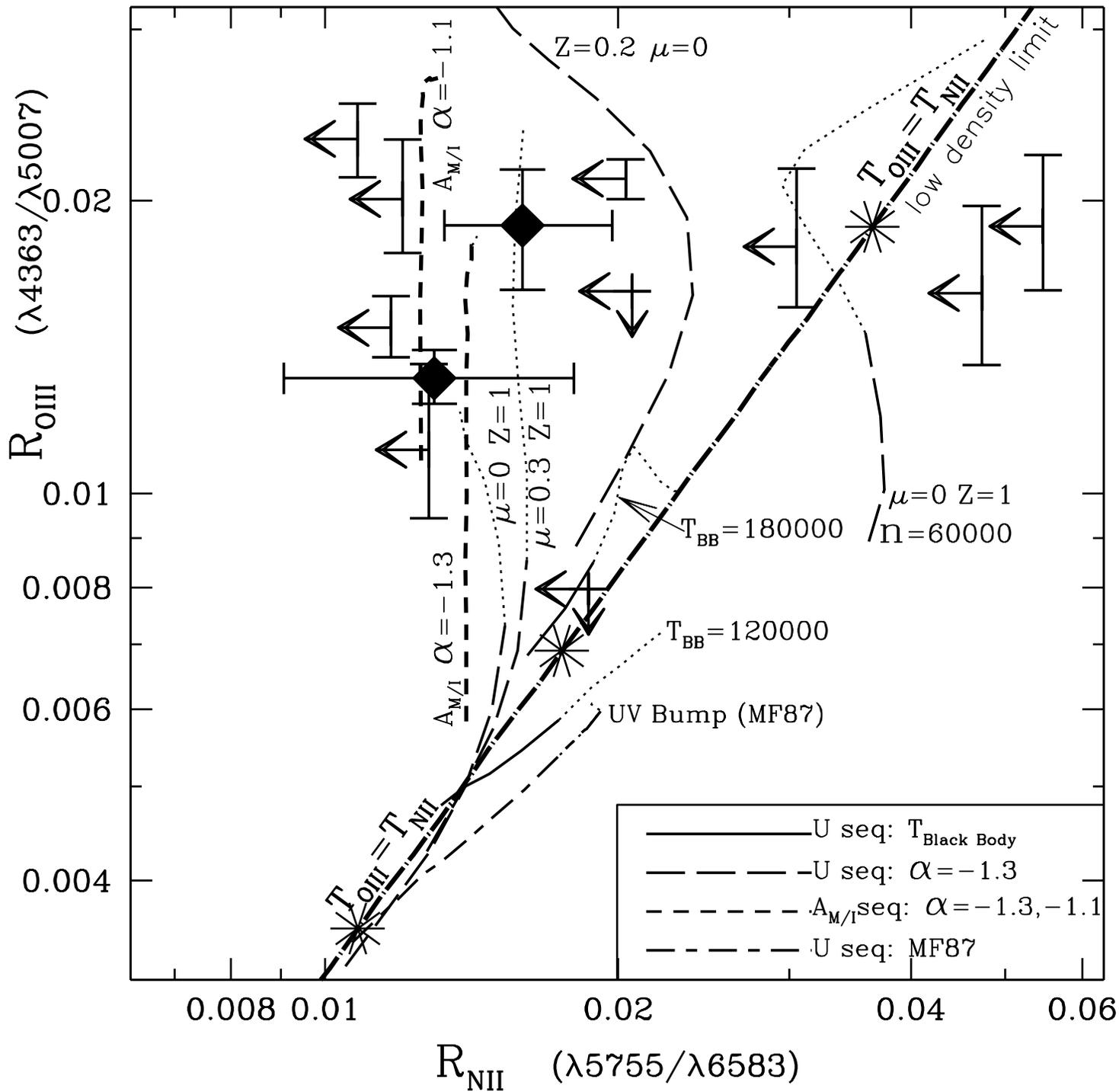
\end{document}